\documentclass[
  reprint,
  superscriptaddress,
  amsmath,amssymb,
  aps,
  prl
]{revtex4-2}

\usepackage{dcolumn}
\usepackage[utf8]{inputenc} 
\usepackage[T1]{fontenc}
\usepackage{helvet} 

\usepackage{dsfont}
\usepackage{mathtools,amssymb,amsmath,bbold,mathrsfs}
\usepackage{mathalfa}
\usepackage{euscript}
\usepackage[bbgreekl]{mathbbol}
\DeclareSymbolFontAlphabet{\mathbbm}{bbold}
\DeclareSymbolFontAlphabet{\mathbb}{AMSb}
\usepackage{bm} 
\usepackage{braket}
\usepackage{slashed}
\usepackage{accents}
\usepackage{csquotes}
\usepackage{upgreek}

\usepackage[table,dvipsnames]{xcolor}
\usepackage{tcolorbox}
\usepackage{enumitem}
\usepackage[colorlinks=true,linkcolor=Blue,citecolor=Blue,linktocpage=true, urlcolor=Blue]{hyperref}
\usepackage{accents}
\usepackage{graphicx} 
\usepackage{adjustbox}
\usepackage[normalem]{ulem}

\DeclareMathAlphabet{\mpala}{T1}{ppl}{m}{it}
\DeclareMathAlphabet{\esstix}{U}{esstixcal}{m}{n}
\DeclareMathAlphabet{\bdx}{U}{BOONDOX-cal}{m}{n}
\DeclareMathAlphabet{\dcal}{U}{dutchcal}{m}{n}
\DeclareMathAlphabet{\bdcal}{U}{dutchcal}{b}{n}
\DeclareMathAlphabet{\bbdx}{U}{BOONDOX-cal}{b}{n}
\DeclareMathAlphabet{\mathpzc}{OT1}{pzc}{b}{n}

\newcommand{\be}{\begin{equation}}
\newcommand{\ee}{\end{equation}}
\newcommand{\bea}{\begin{eqnarray}}
\newcommand{\eea}{\end{eqnarray}}
\newcommand{\bee}{\begin{equation} \begin{aligned}}
\newcommand{\eee}{\end{aligned} \end{equation}}

\newcommand{\ra}{\rangle}

\definecolor{Sepia}{RGB}{244, 232, 201}
\definecolor{paletan}{RGB}{245, 245, 220}
\definecolor{palegray}{RGB}{230,230,230}

\begin{document}


\title{Constraint Variance Identity and\\ the Matter-Geometry Fluctuation Relation}

\author{Yuki Yokokura}
\email{yokokura.yuki@kochi-tech.ac.jp}
\affiliation{School of Engineering Science, Kochi University of Technology, 185, Miyanoguchi, Tosayamada, Kami, Kochi 782-5232, Japan}
\affiliation{RIKEN iTHEMS, Wako, Saitama 351-0198, Japan}

\begin{abstract}
The quantum Hamiltonian constraint vanishes for physical states. For separable Hamiltonians in minisuperspace, its vanishing variance yields a fundamental identity. Under the Born-Oppenheimer approximation, we derive the Matter-Geometry Fluctuation Relation, dynamically linking spacetime clock fluctuations to matter fluctuations. This establishes a quantitative stability condition for the clock against total quantum noise. Finally, we obtain an uncertainty relation between matter energy and relational time. The stable existence of spacetime thus demands quantum fluctuations in matter.
\end{abstract}

\maketitle

\textbf{\textit{Introduction.}}\textemdash 
The universe is a dynamical interplay of spacetime and matter, governed by general relativity. However, understanding its evolution encounters the ``problem of time,'' rooted in the diffeomorphism invariance manifested by the Hamiltonian constraint $H=0$ \cite{Kuchar:1991qf,Isham:1992ms}.
A fundamental approach, conceptually originated from Einstein's ``point coincidences'' \cite{Einstein:1916vd}, is to adopt ``relational time'' \cite{Rovelli:1990jm,Brown:1994py,Rovelli:2001bz,Dittrich:2005kc}; this describes the evolution of the universe in terms of how degrees of freedom change relative to one another, with a monotonically varying variable serving as a physical clock.

In the quantum regime, this approach faces a fundamental challenge. The dynamics are governed by the quantum Hamiltonian constraint, the Wheeler-DeWitt (WDW) equation, $\hat H \vert{}\Psi\ra=0$ \cite{DeWitt:1967yk,Wheeler1968,Kiefer:2004xyv}. Since the clock itself is a quantum entity, it not only intrinsically fluctuates but also suffers from backreaction induced by the quantum fluctuations of other degrees of freedom. In the worst case, the clock becomes too blurred to accurately track time evolution. While conventional relational approaches often assume that such clock fluctuations are negligible \cite{Page:1983uc,Unruh:1989db,Smith:2017pwx,Hoehn:2019fsy,Hoehn:2020epv}, identifying the precise conditions for a stable semiclassical clock remains a major open question. Resolving this requires us to go beyond the first moment (average energy balance) of $\hat H \vert{}\Psi\ra=0$ and evaluate the second moment (energy variance).

In this work, we address this problem in minisuperspace by considering a general separable Hamiltonian $\hat H=\hat H_G(\hat q,\hat p)+ \hat H_M(\hat\phi^a,\hat\pi_a,\hat q)$ with a large number of matter variables ($a=1,2,\dots N\gg1$), using a geometric spacetime variable $q$ as the clock. Since the quantum constraint vanishes for physical states, its variance naturally vanishes as well. However, for separable Hamiltonians, this exact ``zero'' yields a fundamental \textit{Constraint Variance Identity} (CVI), $\Delta H_G^2=\Delta H_M^2$. We employ the Born-Oppenheimer (BO) approximation \cite{Banks:1984cw,Vilenkin:1988yd,Brout:1988ku,Kiefer:1993fg} to explore the physical implications of this identity.

First, we derive the \textit{Matter-Geometry Fluctuation Relation} (MGFR); this dynamically links the quantum fluctuations of the spacetime clock directly to those of the matter sector, governing the total fluctuations in a background spacetime.

Second, the MGFR establishes a quantitative condition for a stable spacetime clock. For statistically independent matter states (e.g., local thermal equilibrium states), the vast number of matter degrees of freedom invokes the central limit theorem, strongly suppressing the fluctuations and stabilizing the clock. For strongly correlated matter states, 
macroscopic quantum fluctuations manifest, destroying the semiclassical clock. As an application, we demonstrate in cosmological models that the clock is stable except for the era near singularities or bounce points.

Finally, as a direct consequence of the MGFR, we derive a \textit{relational time uncertainty relation}. Distinct from the standard time-energy inequality \cite{Mandelstam1945}, which assumes an external time, this relation quantifies the inevitable blurring of time induced by the intrinsic matter fluctuations and their mutual interplay with the clock. Ultimately, it demonstrates that the stable existence of spacetime fundamentally demands intrinsic quantum fluctuations in the matter sector, establishing a new limit in quantum gravity.
\\

\textbf{\textit{Setup.}}\textemdash Let us consider a quantum gravitational system in minisuperspace governed by the quantum Hamiltonian constraint, the WDW equation:
\begin{equation}\label{WDW}
\hat{H}\vert{}\Psi\rangle = 0.
\end{equation}
We assume the Hamiltonian is separable into gravity and matter sectors:
\begin{equation}\label{H}
\hat{H} = \hat{H}_G(\hat{q}, \hat{p}) + \hat{H}_M(\hat{\phi}^a, \hat{\pi}_a, \hat{q}),
\end{equation}
where $[\hat{q}, \hat{p}] = i\hbar$ and $[\hat{\phi}^a,\hat{\pi}_b]=i\hbar \delta^a_b$.
Motivated by the fact that in our universe the number of degrees of freedom of matter fields far exceeds that of the gravitational field, 
we consider a large-$N$ limit for the matter sector ($a=1, 2, \dots, N$ with $N \gg 1$), which universally couples to gravity via $\hat{q}$. As will be clarified in the BO approximation below, we regard the geometric variable $q$ as the physical clock.
\\

\textbf{\textit{Constraint Variance Identity.}}\textemdash
Generally, exploring quantum fluctuations requires evaluating the variance of the relevant observables. To assess the dynamical—rather than merely kinematical—quantum blurriness of the spacetime clock, let us investigate the variance of the dynamical equation itself: the Hamiltonian constraint \footnote{See e.g., \cite{Bojowald:2005cw,Bojowald:2009zzc} for an effective approach using moments.}. For physical states satisfying the constraint \eqref{WDW}, the expectation value evaluated in a well-defined inner product, $\langle\hat H \rangle\equiv \langle \Psi \vert{} \hat{H} \vert{} \Psi \rangle = 0$, trivially yields the average energy balance: $\langle \hat{H}_G \rangle = - \langle \hat{H}_M \rangle$. Subtracting this mean from the original constraint $(\hat H_G+\hat H_M)\vert{}\Psi\rangle=0$, we obtain an operator relation for the fluctuations:
\begin{equation}\label{ope_id}
\delta \hat H_G|\Psi\rangle =-\delta \hat H_M|\Psi\rangle,
\end{equation}
where $\delta \hat H_{G,M} \equiv \hat H_{G,M}- \langle \hat H_{G,M}\rangle$. Taking the norm of this state and using the definition of variance, $\Delta O^2 \equiv \langle (\delta\hat{O})^2 \rangle = \langle \hat{O}^2 \rangle - \langle \hat{O} \rangle^2$, we arrive at the \textit{Constraint Variance Identity} (CVI),
\begin{equation}\label{CVI}
\boxed{\Delta H_G^2 = \Delta H_M^2}~.
\end{equation}

This identity dictates a profound physical reality: the energy variance of gravity ($\Delta H_G^2$) is strictly locked to the energy fluctuations of matter ($\Delta H_M^2$). Importantly, this equality holds even though the two sectors do not commute, $[\hat{H}_G, \hat{H}_M] \neq 0$, due to the universal coupling via $\hat{q}$. Furthermore, calculating directly the total variance $0 = \Delta H^2 = \Delta H_G^2 + \Delta H_M^2 + \langle \delta\hat{H}_G \delta\hat{H}_M + \delta\hat{H}_M \delta\hat{H}_G \rangle$ reveals that the cross term (covariance) emerges but perfectly compensates the individual variances, leading to the CVI \eqref{CVI}. This is a result of the negative correlation arising from the entanglement between the matter and spacetime sectors via the WDW equation.
\\

\textbf{\textit{Born-Oppenheimer Semi-classical Clock.}}\textemdash To employ the geometric variable $q$ as a semiclassical clock and systematically evaluate its quantum fluctuations, the Born-Oppenheimer (BO) approximation provides an appropriate framework \cite{Kiefer:2004xyv,Banks:1984cw,Vilenkin:1988yd,Brout:1988ku,Kiefer:1993fg}. By exploiting the mass-scale separation between the slow heavy gravitational variable $q$ and the fast light matter fields $\phi^a$, we set a wavepacket ansatz:
\begin{align}\label{Psi}
  |\Psi\rangle&=\int dq \psi_G(q)|q\rangle_G \otimes|\psi_M(q)\rangle_M,\\
  \label{psi_G}
    \psi_G(q)&=\frac{1}{(2\pi(\Delta q)^2)^{1/4}}e^{-\frac{(q-\bar q)^2}{4(\Delta q)^2}}e^{\frac{i}{\hbar}S(q)},
\end{align}
where gravity and matter 
are entangled through $q$. 

This is obtained as follows. 
First, expanding the constraint \eqref{WDW} in $\hbar$ with $\vert{}\Psi\rangle=A(q)e^{iS(q)/\hbar}\vert{}q\rangle_G\otimes \vert{}\psi_M(q)\rangle_M$ 
leads to the semiclassical Einstein-Hamilton-Jacobi equation and the relational Schrödinger equation \footnote{See Appendix A for the derivation from a general Hamiltonian $H_G(q,p)$ beyond standard kinetic forms.}:
\begin{align}\label{HJ}
    H_G(q, p=\partial_q S)+ \langle \hat H_M(q)\rangle_M  = 0,\\
    \label{Schrodinger}
    i \hbar\dot q \partial_q \vert{}\psi_M(q)\rangle_M = (\hat{H}_M(q)-\langle \hat H_M(q)\rangle_M) \vert{}\psi_M(q)\rangle_M,
\end{align}
where 
$\hat H_M(q)\equiv H_M(\hat \phi^a, \hat \pi_q,q)$, $\langle \hat H_M(q)\rangle_M\equiv \langle\psi_M(q)|\hat{H}_M(q)|\psi_M(q) \rangle_M$, and $\dot q(q)\equiv \frac{\partial H}{\partial p}|_{q,p=\partial_qS(q)}$. By defining the relational time derivative along the classical trajectory as $\frac{d}{dt} \equiv \dot{q} \partial_q$, \eqref{Schrodinger} reproduces the Schrödinger evolution with a ``time''-dependent Hamiltonian \footnote{Rather than introducing an idealized external clock Hamiltonian, our framework treats the dynamical geometric degree of freedom itself as a physical semiclassical clock. This formulation directly embodies the relational ``point coincidences'' fundamental to general covariance \cite{Einstein:1916vd}, and is valid locally as long as $\dot{q} \neq 0$.}.  
These two equations determine a pair $(S(q),~|\psi_M(q)\rangle_M)$  self-consistently. Then, by superposing such semiclassical solutions, we obtain the wavepacket ansatz \eqref{Psi} peaked around a classical trajectory $\bar q$ with variance $(\Delta q)^2$. 

Importantly, this wavepacket formulation gracefully circumvents the Hilbert space problem of the WDW equation \cite{Kuchar:1991qf,Isham:1992ms}. Since $\vert{}\psi_G(q)\vert{}^2$ is strongly peaked at $\bar q$, the standard kinematical inner product naturally induces a well-defined norm, $\langle\Psi\vert{}\Psi\rangle \approx \langle\psi_M(\bar q)\vert{}\psi_M(\bar q)\rangle_M=1$, where the unitarity is guaranteed by \eqref{Schrodinger}. Furthermore, the expectation values of non-Dirac observables in the matter sector, $\langle \psi_M(q)\vert{}\hat O_M\vert{}\psi_M(q)\rangle_M$, coincide with those of the corresponding relational Dirac observables constructed via group averaging \cite{Hoehn:2019fsy,Hoehn:2020epv}.

Equipped with this physically well-defined and covariant setup, we are now ready to evaluate the quantum fluctuations of the spacetime clock.
\\

\textbf{\textit{Matter-Geometry Fluctuation Relation.}}\textemdash We now apply the BO ansatz \eqref{Psi} to the CVI \eqref{CVI} to derive a dynamical relation between the fluctuations of matter and the clock. Let us evaluate the operator identity \eqref{ope_id} within the BO framework. For the wavepacket \eqref{Psi} with a small $\Delta q$, we have $\langle \hat H_G \rangle\approx H_G(\bar q, \bar p)\equiv \bar H_G$ and $\langle \hat H_M \rangle\approx \langle \psi_M(\bar q)\vert{}\hat H_M(\bar q)\vert{}\psi_M(\bar q)\rangle_M\equiv \bar H_M$, where $\bar p =\partial_q S(\bar q)$. Setting $\hat q=\bar q +\delta \hat q$ and $\hat p=\bar p +\delta \hat p$ and expanding $\hat H_G=H_G(\hat q, \hat p)$ for small $\delta \hat q$ and $\delta \hat p$, we obtain
\begin{align}\label{dH_G}
\delta \hat H_G \approx \frac{\partial \bar H_G}{\partial q}\delta \hat q+\frac{\partial \bar H_G}{\partial p}\delta \hat p.
\end{align}
Similarly, we expand the matter Hamiltonian $\hat H_M=H_M(\hat \phi,\hat \pi,\hat q)\approx H_M(\hat \phi, \hat \pi, \bar q)+\frac{\partial H_M(\hat \phi, \hat \pi, \bar q) }{\partial q}\delta \hat q$. Defining the generalized force operator as
\begin{equation}\label{F_M}
\hat F_M(q)\equiv- \frac{\partial H_M(\hat \phi, \hat \pi, q)}{\partial q},
\end{equation}
we introduce its fluctuation $\delta \hat F_M\equiv\hat F_M(\bar q)-\langle \psi_M(\bar q)\vert{}\hat F_M(\bar q)\vert{}\psi_M(\bar q)\rangle_M$. Noting that $\langle \psi_M(\bar q)\vert{}\frac{\partial H_M(\hat \phi, \hat \pi, \bar q) }{\partial q}\vert{}\psi_M(\bar q)\rangle_M=\frac{\partial \bar H_M(\bar q)}{\partial q}$ \footnote{This can be easily verified using the gauge condition $\langle\psi_M\vert{}\partial_q\psi_M\rangle_M=0$ \cite{Kiefer:1993fg} and the Schrödinger equation \eqref{Schrodinger}.}, we find
\begin{align}\label{dH_M}
\delta \hat H_M\approx\delta \hat H_M^{(0)}+\left(\frac{\partial \bar H_M}{\partial q}-\delta \hat F_M \right)\delta \hat q,
\end{align}
where $\delta \hat H_M^{(0)}\equiv H_M(\hat \phi,\hat \pi,\bar q)-\bar H_M$ represents the pure matter fluctuation operator in the classical background spacetime $\bar q$.
Substituting \eqref{dH_G} and \eqref{dH_M} into the operator identity \eqref{ope_id} and using $\frac{\partial \bar H_G}{\partial p}=\frac{\partial \bar H}{\partial p}$ (with $\frac{\partial  \bar H_M}{\partial p}=0$), we obtain
\begin{align}\label{ope_id_2}
\left(\frac{\partial \bar H}{\partial q}-\delta \hat F_M \right)\delta \hat q+ \frac{\partial \bar H}{\partial p}\delta \hat p=-\delta \hat H_M^{(0)}.
\end{align}

We now take the expectation value of the square of both sides of \eqref{ope_id_2}. Using the property of the wavepacket \eqref{Psi}, the geometric variances evaluate to
\begin{align}\label{dq}
    \Delta q^2&=\langle (\delta \hat q)^2 \rangle=(\Delta q)^2\\
    \label{dp}
    \Delta p^2&=\langle (\delta \hat p)^2 \rangle=(\Delta p_{\rm QM})^2+(\partial_q^2S(\bar q))^2 (\Delta q)^2\\
    \label{dqdp}
    \langle\{\delta \hat q,\delta \hat p \}\rangle&=2 \partial_q^2S(\bar q)(\Delta q)^2,
\end{align}
where $\Delta p_{\rm QM}\equiv \frac{\hbar}{2\Delta q}$ originates from the fundamental quantum-gravitational commutation relation $[\hat q,\hat p]=i\hbar$. For any matter operator $\hat O_M(q)$, we approximate $\langle \hat O_M (q) \rangle\approx \langle \psi_M(\bar q)\vert{}\hat O_M(\bar q)\vert{}\psi_M(\bar q)\rangle_M \equiv \langle \hat O_M\rangle_M$. Furthermore, the cross-terms proportional to $\delta \hat F_M$ vanish (e.g., $\langle \delta \hat F_M (\delta \hat q)^2 \rangle\approx \langle \delta \hat F_M\rangle_M\langle (\delta \hat q)^2 \rangle=0$). This yields the intermediate relation:
\begin{align}
   \left[\left(\frac{\partial \bar H}{\partial q}+\partial_q^2S(\bar q)\frac{\partial \bar H}{\partial p}\right)^2+\langle(\delta \hat F_M)^2\rangle_M\right] \Delta q^2\nonumber\\
   +\left(\frac{\partial \bar H}{\partial p}\right)^2(\Delta p_{\rm QM})^2=\langle (\delta\hat H_M^{(0)})^2 \rangle_M.
\end{align}
Crucially, the Hamilton-Jacobi equation \eqref{HJ} dictates $\bar H(\bar q, \bar p=\partial_q S(\bar q))=0$ along the trajectory, which implies $0=\frac{d\bar H}{dq}=\frac{\partial \bar H}{\partial q}+\partial_q^2S(\bar q)\frac{\partial \bar H}{\partial p}$. The first squared term in the bracket thus identically vanishes. We finally arrive at the \textit{Matter-Geometry Fluctuation Relation} (MGFR):
\begin{equation}\label{MGFR}
\boxed{\left(\frac{\partial \bar H}{\partial p}\right)^2 (\Delta p_{\rm QM})^2=\Delta_0 H_M^2-\Delta F_M^2\Delta q^2}~,
\end{equation}
where $\Delta_0 H_M^2\equiv \langle (\delta \hat H_M^{(0)})^2 \rangle_M$ and $\Delta F_M^2\equiv \langle (\delta \hat F_M)^2 \rangle_M$. Note that this result holds regardless of the details of the separable Hamiltonian \eqref{H}.

This dynamical relation links the spacetime clock and matter at the level of quantum fluctuations. The fluctuation of matter energy ($\Delta_0H_M^2$) forces the momentum of the clock to blur ($\Delta p^2_{QM}$); meanwhile, the spatial fluctuation of the clock ($\Delta q^2$) couples to that of the generalized force ($\Delta F_M^2$), and their mutual fluctuation ($\Delta F_M^2\Delta q^2$) backreacts on the clock. Such a fundamental feedback cycle inevitably arises in quantum gravitational systems. In this sense, the MGFR quantitatively embodies a quantum fluctuation counterpart to Wheeler's maxim \cite{Misner:1973prb}: \textit{quantum matter tells the spacetime clock how to blur, and the fluctuating clock tells quantum matter how to evolve.}

It should be emphasized that the geometric fluctuations $\Delta q$ and $\Delta p_{\rm QM}$ are evaluated within the physical wavepacket \eqref{Psi} solving the  constraint \eqref{WDW}.
Rather than being uncertainties of Dirac observables or unconstrained kinematical variables, they directly quantify the fundamental resolution of the ``partial observable'' \cite{Rovelli:2001bz} serving as a relational clock, which is intrinsically limited by the quantum noise of matter through the MGFR \eqref{MGFR}.

An immediate and striking consequence of \eqref{MGFR} is that a purely classical matter sector without quantum fluctuations ($\Delta_0 H_M^2, \Delta F_M^2 \to 0$) inevitably forces the geometric uncertainty to diverge ($\Delta q = \frac{\hbar}{2\Delta p_{\rm QM}} \to \infty$). This dictates that the stable existence of a semiclassical geometry fundamentally requires intrinsic quantum fluctuations in the matter sector, even when the average background satisfies the classical Einstein equations.
\\

\textbf{\textit{Stability Condition for the Clock.}}\textemdash For the geometric clock to be stable, its quantum fluctuations must be sufficiently smaller than the mean values in both position and momentum:
\begin{equation}\label{cond_qp}
\frac{\Delta q}{|\bar q|}\ll1, \quad \frac{\Delta p_{\rm QM}}{|\bar p|}\ll1,
\end{equation}
where we have approximated the total momentum variance \eqref{dp} by its fundamental quantum bound, $\Delta p^2 \approx \Delta p_{\rm QM}^2$, assuming a suitably prepared wavepacket whose classical phase dispersion is suppressed \footnote{For a sharply peaked wavepacket, the classical phase dispersion $(\partial_q^2 S)^2 \Delta q^2$ is heavily suppressed and negligible compared to the intrinsic uncertainty $\Delta p_{\rm QM}^2$ \cite{Littlejohn:1985ht,Kiefer:1993fg}. See also Appendix B.}. We now derive from the MGFR \eqref{MGFR} a quantitative, dynamical condition to satisfy this stability.

Assuming the fluctuations are small, the mutual term $\Delta F_M^2\Delta q^2$ is sub-leading. The MGFR \eqref{MGFR} reduces to $\left(\frac{\partial \bar H}{\partial p}\right)^2 \Delta p_{\rm QM}^2\approx \Delta_0 H_M^2$. Using the energy balance $\bar H_G=-\bar H_M$ and the relation $\frac{\partial \bar H}{\partial p}=\frac{\partial \bar H_G}{\partial p}$, this leads to
\begin{equation}\label{p_cond}
\boxed{\frac{\Delta p_{\rm QM}}{|\bar p|}=\frac{1}{\eta \sqrt{N_{\rm eff}}}}~,
\end{equation}
where we have introduced
\begin{align}\label{eta_Neff}
\eta\equiv \left|\frac{\bar p}{\bar H_G}\frac{\partial \bar H_G}{\partial p}\right|, \quad \frac{1}{N_{\rm eff}}\equiv\frac{\Delta_0 H_M^2}{(\bar H_M)^2}.
\end{align}
Using $\Delta p_{\rm QM}=\frac{\hbar}{2\Delta q}$, \eqref{p_cond} translates the positional fluctuation bound to
\begin{equation}\label{q_cond}
\boxed{\frac{\Delta q}{|\bar q|}=\frac{\hbar\eta\sqrt{N_{\rm eff}}}{2|\bar q \bar p|}}~.
\end{equation}
Thus, the stability of the clock is dictated by the right-hand sides of \eqref{p_cond} and \eqref{q_cond}.

Let us decode the physical implications. First, noting $\bar p=\partial_qS(\bar q)$, the factor $\eta$ is a function of $\bar q$ that typically takes $\mathcal{O}(1)$ values due to Euler's theorem for homogeneous functions (i.e., $\bar p\frac{\partial \bar H_G}{\partial p}=k\bar H_G$), at least in local regimes (see below for examples). Second, $N_{\rm eff}$ characterizes the relative variance of the matter energy for a large number of fields $N \gg 1$. Since the mean energy scales extensively as $\bar H_M=\mathcal{O}(N)$, the variance scaling depends on the state. For statistically independent matter states, the central limit theorem suppresses the variance to $\Delta_0 H_M^2 = \mathcal{O}(N)$ \cite{Landau:1980mil}, yielding $N_{\rm eff}=\mathcal{O}(N)$. For strongly correlated matter states, macroscopic fluctuations emerge, $\Delta_0 H_M^2 = \mathcal{O}(N^2)$, providing $N_{\rm eff}=\mathcal{O}(1)$.

To satisfy $\Delta p_{\rm QM}/\vert{}\bar p\vert{} \ll 1$ in \eqref{p_cond}, the matter sector must contain macroscopic degrees of freedom and remain statistically independent. If $N$ is small, or if the matter state is strongly correlated, the momentum of the clock becomes fundamentally blurred, breaking down the semiclassical approximation \footnote{In such regimes, $\Delta q$ is no longer small, and the mutual fluctuation term $\Delta F_M^2 \Delta q^2$ in the MGFR \eqref{MGFR} cannot be treated perturbatively.}. In this sense, the stability of the semiclassical spacetime is a macroscopic property anchored by the statistical limit of the matter sector.

While the momentum bound \eqref{p_cond} is governed almost by statistical properties, the positional stability in \eqref{q_cond} explicitly depends on the classical trajectory $\bar p=\bar p(\bar q)$. To fully conclude the stability of the clock, we must evaluate \eqref{q_cond} dynamically within a specific model. 
\\

\textbf{\textit{Example: Cosmological Models.}}\textemdash We apply our bounds to a cosmological model governed by
\begin{equation}\label{cosmology}
\bar H=\bar H_G+\bar H_M=-\frac{\bar p^2}{2\bar q}+\frac{N \epsilon}{\bar q^{3w}} = 0,
\end{equation}
where $q$ represents the scale factor, $\epsilon$ is a constant, and $P=w\rho$ is the equation of state. From \eqref{eta_Neff}, we immediately find $\eta=2$. Using $\bar p=\sqrt{2N\epsilon}\bar q^{-\frac{3w-1}{2}}$, the stability bounds \eqref{p_cond} and \eqref{q_cond} yield
\begin{align}\label{pq_cosmo}
\frac{\Delta p_{\rm QM}}{|\bar p|}=\frac{1}{2 \sqrt{N_{\rm eff}}}, \quad \frac{\Delta q}{|\bar q|}=\frac{\hbar\sqrt{N_{\rm eff}}}{\sqrt{2N \epsilon}}q^{-\frac{3}{2}(1-w)}.
\end{align}
The physical stability thus depends on the scaling of $N_{\rm eff}$ for different cosmic epochs \footnote{See Appendix B for derivations of $N_{\rm eff}$ in each epoch.}.

\noindent (i) \textit{Thermal epoch} ($w=0, 1/3$): When matter is in local thermal equilibrium, the fluctuation-dissipation theorem dictates that the energy variance scales extensively \cite{Landau:1980mil}, giving a constant $N_{\rm eff} = \mathcal{O}(N)$. For $N \gg 1$ and $q \gg 1$, both bounds in \eqref{pq_cosmo} are strongly suppressed. The geometric clock is macroscopically stable.

\noindent (ii) \textit{Inflationary epoch} ($w=-1$): 
$\bar{H}_M\propto \bar q^3$ behaves as a classical cosmological constant. The effective variance is governed by sub-leading quantum field fluctuations on this background, yielding $N_{\rm eff} \sim N \bar q^{3}$. Hence, both in \eqref{pq_cosmo} remain small for $N \gg 1$ and $q \gg 1$. 

\noindent (iii) \textit{Very early epoch} ($w=1$): Modeling matter as $N$ massless free scalar fields in a statistically independent state yields a constant $N_{\rm eff} = \mathcal{O}(N)$. While the momentum fluctuation in \eqref{pq_cosmo} is suppressed by large $N$, the positional fluctuation remains $\mathcal{O}(1)$, 
rendering the clock fragile. The region near the singularity ultimately requires a full quantum-gravitational description.

As another model, consider bouncing cosmologies \cite{Ashtekar:2011ni,Brandenberger:2016vhg}. At the bounce point where the expansion halts ($\dot {\bar q}=\frac{\partial \bar H}{\partial p}= 0$), the factor $\eta$ vanishes. Consequently, the momentum fluctuation \eqref{p_cond} diverges. 
Thus, the semiclassical clock fundamentally breaks down at the bounce, corroborating the fragility seen in case (iii).

These explicit applications demonstrate that our bounds \eqref{p_cond} and \eqref{q_cond} provide a quantitative diagnostic for the breakdown of semiclassical spacetime.
\\

\textbf{\textit{Relational Time Uncertainty Relation.}}\textemdash We discuss a fundamental limit on 
the relational time $t$. Using $\Delta p_{\rm QM}=\frac{\hbar}{2\Delta q}$ and $\dot{\bar q}=\frac{\partial \bar H}{\partial p}$, we rewrite the MGFR \eqref{MGFR} as
\begin{equation}\label{uncertainty}
\boxed{\Delta_0 H_M \Delta t =\frac{\hbar}{2} \Gamma\geq\frac{\hbar}{2}}~,
\end{equation}
where $\Delta_0 H_M\equiv\sqrt{\Delta_0 H_M^2}$ and
\begin{align}\label{dt}
\Delta t \equiv \frac{\Delta q}{|\dot{\bar q}|}, \quad \Gamma\equiv\left(1-\frac{\Delta F_M^2 \Delta q^2}{\Delta_0 H_M^2}\right)^{-\frac{1}{2}}\geq1~.
\end{align}
Here, $\Delta t$ characterizes the timescale required for the clock to resolve a geometric change, representing the intrinsic blurriness of the relational time $t$.

This \textit{Relational Time Uncertainty Relation} profoundly differs from the standard inequality $\Delta E \Delta t_{\rm ext}\geq \frac{\hbar}{2}$ \cite{Mandelstam1945}. In the latter, $t_{\rm ext}$ is an ideal external parameter, and the bound originates from the matter Schrödinger evolution. In contrast, the obtained bound \eqref{uncertainty} is governed by the relational time $t$, and its $\hbar$ originates purely from the quantum-gravitational non-commutativity $[\hat q,\hat p]=i\hbar$.

Crucially, the trade-off $\Delta_0 H_M \Delta t \geq \frac{\hbar}{2}$ dictates that taking a perfect-clock limit ($\Delta t\to 0$) forces the matter energy fluctuation to diverge ($\Delta_0 H_M \to \infty$). Thus, neither a strictly sharp geometry nor perfectly coherent matter can stably exist; both must inherently fluctuate to sustain a semiclassical universe. Furthermore, the correction $\Gamma \geq 1$, stemming from the mutual fluctuation $\Delta F_M^2 \Delta q^2$, degrades the temporal resolution beyond the standard quantum limit. This uncertainty should play a foundational role in defining the temporal resolution of any quantum-gravitational system.
\\

\textbf{\textit{Conclusion.}}\textemdash 
By investigating the variance of the Wheeler-DeWitt equation, we derived the constraint variance identity and captured the dynamical fluctuations of quantum gravity. Within the Born-Oppenheimer framework, this identity yields the matter-geometry fluctuation relation and the relational time uncertainty relation. These serve a profound dual purpose: they provide a quantitative diagnostic for the stability of semiclassical spacetime, and they imply a novel physical view. Namely, the macroscopic stability of our universe—comprising spacetime and matter—is sustained precisely \textit{because} it intrinsically fluctuates.

This fluctuation framework opens several important avenues. Extending this approach to multi-variable minisuperspace settings, such as spherically symmetric static spacetimes, is a natural next step. Further generalization to full quantum field theory promises deeper insights, though it will require careful regularization of composite operators. Furthermore, our quantitative diagnostic should offer a robust foundation to explore novel quantum-gravitational phenomenology. Ultimately, evaluating the exact constraint variance identity without the Born-Oppenheimer approximation may unveil the non-perturbative structure of quantum spacetime-matter entanglement, where the inevitable blurriness of spacetime alters the fabric of the universe.
\\


\textit{Acknowledgments}\textemdash I would like to thank Lucien Hardy and Masato Itami for the insightful discussions, and I am also grateful for the fruitful discussions during the workshop \textit{Relativity Reframed} at Perimeter Institute.
\\
\bibliography{biblio}

\clearpage
\onecolumngrid
\begin{center}\bfseries Appendix\end{center}
\twocolumngrid

\section*{A: Effective Relational Dynamics from a General Hamiltonian}\label{App:Emergence}

In typical derivations of the relational Schrödinger equation from the WDW equation, the gravitational Hamiltonian is restricted to a standard kinetic form, $H_G = -\frac{p^2}{2M}+ V(q)$ \cite{Kiefer:2004xyv}. Here, we show that the Hamilton-Jacobi equation \eqref{HJ} and the Schrödinger equation \eqref{Schrodinger} can be derived from the WDW equation with a \textit{general} Hamiltonian $H_G(q, p)$, where $\frac{\partial H_G}{\partial p}$ may have a nonlinear dependence on $p$. This universality ensures that our results can be extended to non-perturbative effective \cite{Bojowald:2005cw,Corichi:2007tf,Bojowald:2009zzc,Ashtekar:2011ni}.

We solve the general WDW equation $(\hat{H}_G + \hat{H}_M)|\Psi\rangle = 0$ using the Born-Oppenheimer ansatz:
\begin{equation}\label{Psi_ansatz}
    |\Psi\rangle = A(q) e^{iS(q)/\hbar} |\psi_M(q)\rangle_M~,
\end{equation}
where $A(q)$ and $S(q)$ are real functions, and the matter state $|\psi_M(q)\rangle_M$ is chosen to satisfy the standard Berry gauge condition $\langle \psi_M | \partial_q \psi_M \rangle_M = 0$ \cite{Kiefer:2004xyv}. To evaluate the action of the general geometric operator $\hat{H}_G(\hat{q}, \hat{p})$ with $\hat{p} = -i\hbar\partial_q$, we expand it up to $\mathcal{O}(\hbar)$ around the classical momentum $p_0(q) = \partial_q S(q)$ by using the Weyl ordering prescription \cite{Littlejohn:1985ht}:
\begin{equation}
    \hat{H}_G(q,\hat p) \approx H_G(q,p_0) + \frac{1}{2}\left[F(q)(\hat p-p_0) + (\hat p-p_0) F(q)\right],
\end{equation}
where $F(q) \equiv \frac{\partial H_G}{\partial p}(q,p_0(q))$. Applying this to \eqref{Psi_ansatz} yields 
\begin{align}
    \hat H_G |\Psi\rangle &= H_G(q,p_0) Ae^{iS/\hbar}|\psi_M\rangle_M \nonumber\\
    &\quad + (-i\hbar)e^{iS/\hbar}\left[\frac{1}{2A}\partial_q(A^2F)|\psi_M\rangle_M + AF\partial_q|\psi_M\rangle_M \right].
\end{align}
Introducing the velocity $\dot{q}(q) \equiv \frac{\partial H_G}{\partial p}(q,\partial_qS(q)) = F(q)$, and dividing the overall equation by $A e^{iS/\hbar}$, the WDW equation reduces, up to $\mathcal{O}(\hbar)$, to  
\begin{align}\label{WDW_semi}
    0 &= \left(H_G(q,p_0) + \hat H_M(q)\right) |\psi_M(q)\rangle_M \nonumber\\
    &\quad + (-i\hbar)\left[\frac{1}{2A^2}\partial_q(A^2\dot q)|\psi_M\rangle_M + \dot q\partial_q|\psi_M\rangle_M \right].
\end{align}

Next, we separate this into two parts. 
Multiplying ${}_M\langle \psi_M |$ from the left, using the gauge condition, and defining the mean matter energy as $\bar{H}_M(q) \equiv {}_M\langle \psi_M(q) | \hat{H}_M(q) | \psi_M(q) \rangle_M$, the real part $\mathcal{O}(\hbar^0)$ and the imaginary part $\mathcal{O}(\hbar^1)$ yield, respectively,
\begin{equation}
    H_G(q, \partial_q S) + \bar{H}_M(q)= 0, \quad \partial_q \left( A^2 \dot{q} \right)= 0.
\end{equation}
The first equation is the semi-classical Einstein-Hamilton-Jacobi equation \eqref{HJ}, and the latter represents the probability conservation law along the trajectory. 
Finally, substituting these relations back into the unprojected equation \eqref{WDW_semi}, we arrive at the relational Schrödinger equation \eqref{Schrodinger}: 
\begin{equation}
    i\hbar \dot{q} \partial_q |\psi_M(q)\rangle_M = \left( \hat{H}_M(q) - \bar{H}_M(q) \right) |\psi_M(q)\rangle_M~.
\end{equation}
This demonstrates that self-consistent interaction between the matter state and a general semi-classical geometric degree of freedom is sufficient for the dynamical realization of relational time in a sector where $\dot q\neq0$.

\section{B: Evaluation of $N_{\rm eff}$ in the Cosmological Model}\label{App:Neff}
We here provide the detailed evaluation of  $N_{\rm eff} \equiv \frac{\bar{H}_M^2}{\Delta_0 H_M^2}$ for each epoch of the the cosmological model \eqref{cosmology}. 
Before proceeding, we verify that $\frac{\Delta p}{|\bar p|}\approx \frac{\Delta p_{\rm QM}}{|\bar p|}$ holds as long as $\frac{\Delta q}{|\bar q|}\ll1$. Noting $\partial_q^2 S(\bar q)=\partial_q \bar p$ and $\bar p=\sqrt{2N\epsilon}\bar q^{-\frac{3w-1}{2}}$, we can use \eqref{dp} and show 
\begin{align}\label{dp_dq_cos}
    \frac{\Delta p^2}{|\bar p|^2}&=\frac{(\Delta p_{\rm QM})^2}{|\bar p|^2}+(\bar q \partial_q \log \bar p)^2 \frac{\Delta q^2}{|\bar q|^2}\nonumber\\
    &=\frac{(\Delta p_{\rm QM})^2}{|\bar p|^2}+\left(\frac{3w-1}{2}\right)^2 \frac{\Delta q^2}{|\bar q|^2}\approx \frac{(\Delta p_{\rm QM})^2}{|\bar p|^2}.
\end{align}
\\

\noindent \textbf{(i) Thermal Epoch} ($w=0, 1/3$).
When the matter sector is in local thermal equilibrium, macroscopic quantities can be evaluated using thermodynamics. First, the fluctuation-dissipation theorem relates the energy variance in a thermal state at temperature $T$, $\Delta_0 H^2_M$, to the heat capacity at constant volume, $C_V\equiv \left(\frac{\partial E}{\partial T}\right)_V$, as \cite{Landau:1980mil}
\begin{equation}\label{FDT}
\Delta_0 H^2_M=k_B T^2 C_V.
\end{equation}
Setting $E=\bar H_M=\rho V$ and applying $P=w\rho$ to a thermodynamic formula $\left(\frac{\partial E}{\partial V}\right)_T=T\left(\frac{\partial P}{\partial T}\right)_V-P$ \cite{Landau:1980mil}, we find 
\begin{equation}\label{rho_w}
    \rho=A_w T^{\frac{1+w}{w}},
\end{equation}
where $A_w$ is a constant dependent on $w$.
Then, we have $C_V=V\frac{\partial \rho}{\partial T}=\frac{1+w}{w}\frac{\bar H_M}{T}$, and \eqref{FDT} becomes $\Delta_0 H^2_M=\frac{1+w}{w}k_B T \bar H_M$, leading to 
\begin{equation}\label{Neff_thermal1}
    N_{\rm eff}\equiv\frac{\bar H_M^2}{\Delta_0H_M^2}=\frac{w}{1+w}\frac{\bar H_M}{k_B T},
\end{equation}
which represents a purely thermodynamic relation.

Now, we introduce the geometric expansion. Setting $V=v_0 \bar q^3$ and  $E=\bar{H}_M = \frac{N \epsilon}{\bar q^{3w}}$ in $\rho=\frac{E}{V}$ and equating it to \eqref{rho_w}, we can obtain $k_BT=\tau_w \bar q^{-3w}$, where $\tau_w$ is another constant. 
Substituting this and $\bar{H}_M = \frac{N \epsilon}{\bar q^{3w}}$ into \eqref{Neff_thermal1}, we arrive at 
\begin{equation}
    N_{\rm eff}=\frac{w}{1+w}\frac{\epsilon}{\tau_w}N\equiv c_w N,
\end{equation}
where $c_w$ is a $\mathcal{O}(1)$ dimensionless constant. Thus, in the thermal epoch, $N_{\rm eff}$ is a constant proportional to $N$.
\\

\noindent \textbf{(ii) Inflationary Epoch} ($w=-1$). During inflation, the macroscopic background is driven by a vacuum energy (cosmological constant $\Lambda=N \epsilon$) such that $\bar{H}_M= \Lambda \bar q^3$, which does not fluctuate. The physical matter sector consists of $N$ free massless scalar fields $\phi_a$ ($a=1,\dots,N$) fluctuating on this de Sitter background.

Assuming all scalar fields share the Bunch-Davies (BD) vacuum $\vert{}0\rangle_{\rm BD}$, cross-field correlations vanish. Thus, the matter Hamiltonian variance in a physical volume $V \propto q^3$ reduces to $N$ times the integrated two-point correlator of a single field's normal-ordered energy density $:\hat \rho^{(1)}:$, 
\begin{equation}
\Delta_0 H_M^2 = N \int_V d^3x \int_V d^3y \langle 0| :\hat\rho^{(1)}(x): :\hat \rho^{(1)}(y): |0 \rangle_{\rm BD}.
\end{equation}
In the BD vacuum, this correlator decays rapidly for physical distances beyond the Hubble radius $H^{-1}$. A macroscopic volume $V \gg H^{-3}$ thus acts as an ensemble of $\mathcal{N}_{\rm Hubble} = V H^3$ statistically independent domains. This cluster decomposition, together with spatial translation invariance, reduces the double integral to a volume-law scaling:
\begin{equation}
\Delta_0 H_M^2 \simeq N V \int_V d^3z \langle0|:\hat\rho^{(1)}(0): :\hat \rho^{(1)}(z): |0 \rangle_{\rm BD} \sim N V H^5~.
\end{equation}
Using $\bar H_M=N V \langle0| \hat\rho^{(1)} |0\rangle_{\rm BD}\sim NVH^4$, we find
\begin{equation}
N_{\rm eff} = \frac{\bar{H}_M^2}{\Delta_0 H_M^2} \sim \frac{(N V H^4)^2}{N V H^5} = N V H^3 = N \mathcal{N}_{\rm Hubble} \propto N q^3~.
\end{equation}
Dictated by the law of large numbers, the relative fluctuations decay as $(N\mathcal{N}_{\rm Hubble})^{-1/2}$. This indicates that $N_{\rm eff}$ grows explosively as the expanding universe accumulates independent Hubble volumes, robustly ensuring the stability of the geometric clock.
\\

\noindent \textbf{(iii) Very Early Epoch} ($w=1$). 
The matter sector for $w=1$ is described by stiff matter, which can be microscopically modeled by $N$ massless free scalar fields with the Hamiltonian $\hat{H}_M (q) = \sum_{a=1}^N \hat{H}_a(q)$, where $\hat{H}_a(q) = \frac{\hat \pi_a^2}{2q^3}$. In the deep pre-inflationary regime near the initial singularity, kinetic energy strictly dominates over any interaction potentials. Absent mechanisms to generate quantum entanglement among different species, it is reasonable to assume that the matter state $\vert{}\psi_M(q)\rangle_M$ is uncorrelated in the asymptotic limit $q \to 0$, where the $N$ scalar fields are statistically independent:
\begin{equation}
|\psi_M\rangle_M = |\phi_1\rangle \otimes |\phi_2\rangle \otimes \cdots \otimes |\phi_N\rangle~.
\end{equation}
Because $\langle \hat{H}_a \hat{H}_b \rangle_M = \langle \hat{H}_a \rangle_a \langle \hat{H}_b \rangle_b$ holds for $a \neq b$, the total variance is the sum of the individual variances:
\begin{equation}
\Delta_0 H_M^2 = \sum_{a=1}^N \Delta_0 H_a^2~.
\end{equation}
Since $\hat{H}_a(q) \propto q^{-3}$, the intrinsic variance of a single field scales as $\Delta_0 H_a^2 = \frac{\sigma^2}{q^6}$ for some constant $\sigma^2$ determined by each state $\vert{}\phi_a\rangle$. Therefore, the total variance is $\Delta_0 H_M^2 = N \frac{\sigma^2}{q^6}$.
Using $\bar H_M=\frac{N\epsilon}{q^3}$, we obtain 
\begin{equation}
N_{\rm eff} = \frac{(N \epsilon / q^3)^2}{N \sigma^2 / q^6} = \left(\frac{\epsilon^2}{\sigma^2}\right) N = \mathcal{O}(N)~.
\end{equation}
Unlike the inflationary epoch, 
it does not gain any volume-law enhancement from the scale factor $q$. As shown in the main text, this leads to an $\mathcal{O}(1)$ relative positional fluctuation $\Delta q / \vert{}\bar{q}\vert{}$, fundamentally breaking the stability of the clock.

\end{document}